\documentclass[aps,pr,reprint,superscriptaddress]{revtex4-1}
\usepackage{graphicx}
\usepackage{dcolumn}
\usepackage{bm}
\usepackage{color}
\usepackage{xcolor}
\usepackage{multirow}
\usepackage{hyperref}
\usepackage{amssymb}
\usepackage{color}
\usepackage{graphics}\usepackage{epsfig}
\usepackage{footnote}
\usepackage{amsmath}
\usepackage{multirow}
\usepackage{graphicx}
\usepackage{tabularx}
\usepackage[normalem]{ulem}

\definecolor{darkgreen}{rgb}{0,0.6,0}

\begin{document}
\title{Antiferromagnetic stripe phase and large-gap insulating ground state of the correlated $\sqrt{3}\times\sqrt{3}$~R30$^{\circ}$-Sn/Si(111) single atomic layer}
\author{M.~Torkzadeh}
\thanks{Both M. Torkzadeh and M. Iannetti contributed equally to this work, with experimental and theoretical contributions respecti vely, and are therefore considered co–first authors.}
\affiliation{Sorbonne Universit\'e, CNRS, Institut des Nanosciences de Paris, UMR7588, F-75252 Paris, France}
\author{M.~Iannetti}
\thanks{Both M. Torkzadeh and M. Iannetti contributed equally to this work, with experimental and theoretical contributions respectively, and are therefore considered co–first authors.}
\affiliation{Department of Physical and Chemical Sciences, University of L'Aquila, Via Vetoio 10, I-67100 L'Aquila (Italy)}
\affiliation{CNR-SPIN c/o Department of Physical and Chemical Sciences, University of L'Aquila, Via Vetoio 10, I-67100 L'Aquila (Italy)}
\author{M.~Liz\'ee}
\affiliation{Sorbonne Universit\'e, CNRS, Institut des Nanosciences de Paris, UMR7588, F-75252 Paris, France}
\author{A.~Thakur}
\author{M.~Herv\'e}
\author{F.~Debontridder}
\author{P.~David}
\affiliation{Sorbonne Universit\'e, CNRS, Institut des Nanosciences de Paris, UMR7588, F-75252 Paris, France}
\author{M.~Casula}
\affiliation{Institut de Min\'eralogie, de Physique des Mat\'eriaux et de Cosmochimie (IMPMC), Sorbonne Universit\'e, CNRS UMR 7590, MNHN, 4 Place Jussieu, 75252 Paris CEDEX 05, France}
\author{G.~Profeta}
\affiliation{Department of Physical and Chemical Sciences, University of L'Aquila, Via Vetoio 10, I-67100 L'Aquila (Italy)}
\affiliation{CNR-SPIN c/o Department of Physical and Chemical Sciences, University of L'Aquila, Via Vetoio 10, I-67100 L'Aquila (Italy)}
\author{T.~Cren}
\affiliation{Sorbonne Universit\'e, CNRS, Institut des Nanosciences de Paris, UMR7588, F-75252 Paris, France}
\author{M.~Calandra}
\affiliation{Sorbonne Universit\'e, CNRS, Institut des Nanosciences de Paris, UMR7588, F-75252 Paris, France}
\affiliation{Dipartimento di Fisica, Università di Trento, via Sommarive 14, I-38123 Povo, Italy}
\author{C.~Tresca}
\email{cesare.tresca@spin.cnr.it}
\affiliation{CNR-SPIN c/o Department of Physical and Chemical Sciences, University of L'Aquila, Via Vetoio 10, I-67100 L'Aquila (Italy)}
\author{C.~Brun}
\email{christophe.brun@sorbonne-universite.fr}
\affiliation{Sorbonne Universit\'e, CNRS, Institut des Nanosciences de Paris, UMR7588, F-75252 Paris, France}


\begin{abstract}
The 1/3 monolayer Sn layer on Si(111) has long been considered a benchmark system for exploring two-dimensional Mott physics, owing to its narrow bandwidth and sizable on-site Coulomb repulsion. Previous experiments suggested the emergence of a low-temperature Mott insulating phase with an energy gap of only a few tens of meV, while theory predicted a possible antiferromagnetic ordering that remained experimentally elusive.
Here, by combining low-temperature scanning tunneling microscopy/spectroscopy with first-principles calculations, we reveal that the $\sqrt{3}\times\sqrt{3}$~R30$^{\circ}$-Sn/Si(111) surface undergoes a transition below 30 K into a robust insulating state characterized by a remarkably large gap of about 440 $\pm$ 120~meV at 4~K, five to ten times larger than previously reported. Quasiparticle interference imaging uncovers a well-defined $2\sqrt{3}\times\sqrt{3}$~R30$^{\circ}$-Sn/Si(111) superstructure, providing direct evidence for a two-dimensional stripe-like antiferromagnetic order. Ab initio calculations reveal that the Si substrate stabilizes this phase through strong nonlocal Sn–Sn interactions, highlighting the decisive role of substrate-driven correlations in the $\sqrt{3}\times\sqrt{3}$~R30$^{\circ}$-Sn/Si(111) system.
\end{abstract}

\maketitle

\section{Introduction}
\label{intro}
Since the discovery of high-temperature superconductivity in cuprates materials, many efforts were made to single out their common underlying aspects. A general attempt was to relate these common features to the properties of their parent undoped compounds which are Mott insulators. A reasonable consensus is that unconventional d-wave superconductivity is induced from electron/hole (e/h) doping of a Mott insulating state \cite{Lee2006}. Nevertheless, the chemical complexity of these multilayered materials is so high that several orders compete with each other. Thus, disentangling these competing orders using simpler and possibly better characterized systems is highly desirable, starting from a simpler paradigmatic Mott insulator. Following this idea, the search for true simple two-dimensional (2D) Mott insulating materials seems relevant and appealing.

In this context, particular attention has been drawn to a class of 2D surface crystals believed to be a prototypical realization of a Mott system on a triangular lattice. They consist of single-layer of metal atoms (Sn or Pb) grown on semiconducting substrates like Si(111) or Ge(111) at different coverage level \cite{Carpinelli1996,Santoro,PhysRevLett.98.086401,Hansmann2013a,Iannetti_2025}. In the high-temperature phase, the 2D surface atoms form a triangular $\sqrt{3}\times\sqrt{3}$~R30$^{\circ}$ reconstruction with respect to the Si(111) surface. Each tetravalent metal atom provides a free electron leading to a single half-filled electronic band confined at the surface and isolated from bulk bands. Due to the large nearest-neighbor distance between Sn or Pb atoms, this band is very narrow and many-body effects become significant.
Most low-energy model Hamiltonians concentrate on the on-site Coulomb term $U$ under the assumption that it dominates the interaction landscape \cite{Li2011,PhysRevLett.98.086401,Tresca2018,Tresca2021,Tresca2023}. Nonetheless, several theoretical studies have shown that non-local Coulomb interactions ($V$) can play a critical role in determining the electronic ground state and phase diagram \cite{Hansmann2013a,cao_prb2018,Iannetti_2025}. The magnitudes of $U$ and $V$ depend on the adatom species and environment \cite{schuwalow2010,Hansmann2013a}, and their competition gives rise to a rich set of correlated phases.

\begin{figure*}
\centering
\includegraphics[width=\linewidth]{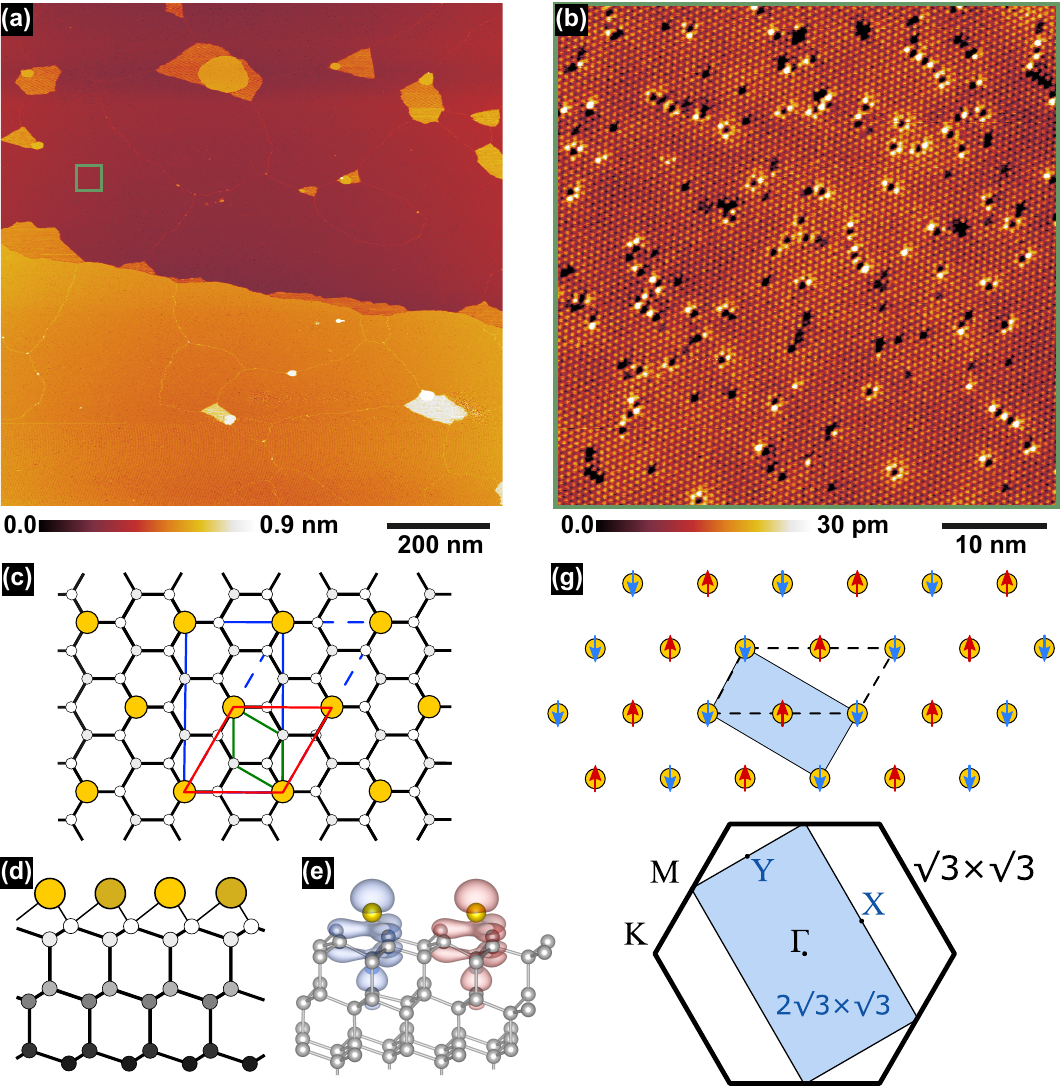}
\caption{(color online) (a) Large scale STM topography of the $\sqrt{3}\times\sqrt{3}$~R30$^{\circ}$-Sn/Si(111) sample measured with $V_{bias}=-1.0~$V and  $I=20~$pA. (b) Small scale area measured in the single $\sqrt{3}\times\sqrt{3}$ domain indicated by the yellow square in panel (a). All measurements were carried out at $T= 4.2$~K. (c) Top view of the $\sqrt{3}\times\sqrt{3}$~R30$^{\circ}$ and 2$\sqrt{3}\times\sqrt{3}$~R30$^{\circ}$-Sn/Si(111) structures (red and blue respectively) where yellow balls represent top Sn atoms, Si atoms are reported in gray scale from light (topmost layer) to dark (inner layer). The elementary Si(111) surface unit cell is shown in green. (d) Corresponding side view projected on the dashed line marked in panel (c). (e) Electronic wavefunctions at the $\Gamma$ point calculated in the HSE06-DFT framework at an energy closest to $\varepsilon_F$: the strong hybridization of Sn orbitals with neighboring Si orbitals leads to a spatial vertical extension up to the third Si(111) plane. g) Proposed magnetic ordering of the spin $1/2$ lattice at low temperature. The relative Brillouin zones are indicated.}\label{topos_structure}
\end{figure*}

In the following we focus on the more studied $\sqrt{3}\times\sqrt{3}$~R30$^{\circ}$-Sn/Si(111) ($\sqrt{3}$-Sn). Structure-sensitive measurements reported the absence of structural transition at helium temperature \cite{Morikawa}. Angular-resolved photoemission (ARPES) results \cite{Lobo2003,ncomms2617,Iannetti_2025,Hirahara2025} and scanning tunneling microscopy/spectroscopy (STM/STS) studies \cite{Modesti2007,Ming2017,Iannetti_2025} showed a strong reduction of density of states (DOS) at the Fermi level between room temperature and 30 to 40~K, typically by nearly a factor of ten, accompanied by a transfer of spectral weight to lower binding energies. Nevertheless the DOS was found to remain finite at the Fermi energy ($\varepsilon_F$) in all these works. To the best of our knowledge, only two works reported spectroscopic results below 30~K at helium temperature by STS \cite{Modesti2007,Odobescu2017}, suggesting the possible opening of a true energy gap ($\Delta$) of less than 100~meV at  $\varepsilon_F$. Regarding the possible spin ordering, a spin-resolved ARPES study performed at 40~K reported the measurement of a very small spin polarization consistent with a collinear stripe antiferromagnetic order of periodicity $2\sqrt{3}\times\sqrt{3}$~R30$^{\circ}$, while the partial gap opening can be estimated in the energy range 200-250~meV \cite{Jaeger2018}. However, the complexity induced by the presence of three possible orientional domains together with the $\mu$m scale of the ARPES spot measurements request for further clarification.
Due to the conflicting  low-temperature spectral features and the lack of a precise value of the energy gap it is difficult to draw a clear picture of the groundstate and its magnetic properties. This results in the absence of a unique comprehensive minimal model able to capture the magnetic ground state. This limitation becomes particularly evident in light of the emergence of an unconventional, possibly topological superconducting phase upon hole doping the $\sqrt{3}$-Sn\cite{Ming2017,Wu2020,Ming2023,Weitering_arxiv}, when trying to relate current theoretical descriptions \cite{Hansmann2013a,cao_prb2018,Wolf2022,Iannetti2026,DiSante_arxiv} and the observed phenomenology. The absence of a reliable microscopic picture of the $\sqrt{3}$-Sn system questions  the origin of superconductivity. A consistent theoretical framework is therefore indispensable to place Sn/Si(111) on a solid microscopic footing.

In the present work, we resolve the elusive nature of the $\sqrt{3}$-Sn ground state. Using high-resolution spatial and spectral STM/STS measurements, we show that below 30~K an insulating state develops with a large gap $\Delta$ of about 440 $\pm$ 120~meV at 4~K, a value five to ten times larger than previously suggested both experimentally and theoretically. A signature of a magnetic order at 4~K is revealed using quasi-particle interferences in single structural domains. The magnetic order gives rise to a sharp non-dispersive superstructure associated to a 2$\sqrt{3}\times\sqrt{3}$~R30$^{\circ}$ period, consistent with the  occurrence of a 2$\sqrt{3}\times\sqrt{3}$~R30$^{\circ}$ collinear antiferromagnetic order. These experimental results are fully supported by density functional theory (DFT) calculations using hybrid functionals, for both the energy gap size and the nature of the magnetic state. In addition, our results indicate that the insulating antiferromagnetic state in this system cannot be fully captured adequately by a purely local Mott-Hubbard description. Instead, substrate-mediated non-local interactions arising from Sn wavefunction hybridization with the silicon substrate play a decisive role and must be included both in one-band models and first-principles calculations to achieve quantitative agreement with experiments, similarly to what happens in the case of Pb/Ge(111)\cite{Tresca2021}. This makes $\sqrt{3}$-Sn a paradigmatic platform for studying correlated 2D electron systems in which non-local Coulomb interaction and substrate effects control the emergent electronic and magnetic order.

\section{Experimental results}
\label{sec_exp}

The $\sqrt{3}$-Sn single-layer phase was prepared following established procedures \cite{ncomms2617,Ming2017,Odobescu2017}; details of the growth and post-growth optimization are given in Appendix~\ref{APPENDIX_experiment}. Careful optimization yields large, structurally clean $\sqrt{3}$-Sn domains (Fig.~\ref{topos_structure}). Typical single-domain lateral sizes exceed 100 nm and are bounded by atomically sharp, curved domain boundaries. Small surface regions occasionally exhibit a denser $2\sqrt{3}\times2\sqrt{3}$~R30$^\circ$ phase, located either within terraces or near step edges, consistent with previous reports \cite{PhysRevB.97.195402}.

\begin{figure*}
\centering 
{\includegraphics[width=2\columnwidth]{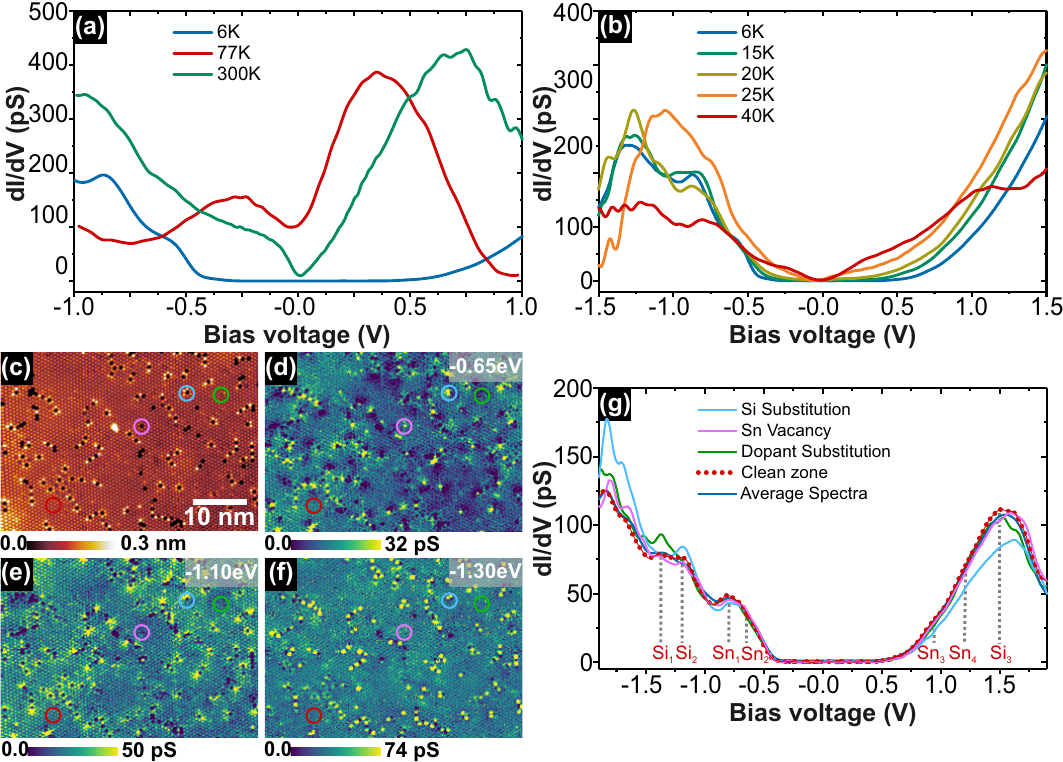}}
\caption{(color online) Temperature dependent $dI/dV(E=eV_{bias})$ spectra acquired by scanning tunneling spectroscopy in single $\sqrt{3}\times\sqrt{3}$~R30$^\circ$-Sn/Si(111) domains. (a) Comparison between tunneling spectra acquired at room temperature (RT), 77~K and 4~K. A large energy gap has set in at 4~K. These spectra are measured between [-1;1]~V using a set-point at $I=200$~pA for $V_{bias}$=-1~V. (b) Temperature evolution of the $dI/dV(E)$ tunneling spectra between $37.6$~K and 6~K. The energy gap opening starts around ~25-30K. These spectra are measured between [-2;2]~V using a set-point at $I=200$~pA for $V_{bias}$=-2~V. (c) STM topography with atomic resolution measured in a $50\times50$~nm$^2$ area at $T=4.2$~K in $\sqrt{3}\times\sqrt{3}$~R30$^{\circ}$-Sn/Si(111) with $V_{bias}=-2.0~$V and  $I=50~$pA. (d,e,f) $dI/dV(E,x,y)$ spectroscopic map recorded at $E=-1.30$~eV, $E=-1.10$~eV, $E=-0.65$~eV extracted from a $256\times256$ grid of spectroscopic maps measured in the energy range [-2;+2]~eV. The Fourier transform of such $dI/dV(E,x,y)$ maps gives the QPI features presented in Fig.~\ref{QPI_mag_order}. (g) Average spectrum of the whole $50\times50$~nm$^2$ area and individual spectra acquired above three typical defects indicated by the corresponding circles in images c-f: Si-substitution, Sn-vacancy and dopant-substitution, and in a location free of defects (labelled clean zone). We marked with Sn$_i$ and Si$_i$ the corresponding main tin and silicon states peaks in the density of states (see text).
}\label{comp_dI_dV}
\end{figure*}

Scanning tunneling microscopy and spectroscopy (STM/STS) measurements were carried out on well-identified $\sqrt{3}$-Sn single domains having a lateral size larger than 50-100~nm using metallic PtIr tips. Representative tunneling conductance spectra are shown in the panel a) of Fig.~\ref{comp_dI_dV} at room temperature (red), 77~K (green) and 4~K (blue). The room temperature spectrum is in good agreement with previous STS\cite{Modesti2007}, ARPES\cite{Modesti2007,Lobo2003,Hirahara2025} and inverse photoemission results \cite{Charrier2001}. It features LDOS maxima around $-0.3$~eV and $+0.35$~eV and suggests an experimental bandwidth larger than $1$~eV. The strong modification of the spectrum seen at 77~K also agrees with the large changes reported by STS \cite{Modesti2007,Ming2017} or ARPES \cite{Modesti2007,ncomms2617,Jaeger2018} results. The spectral weight across most of the bandwidth is depleted and appears to be transferred to lower binding energies and larger unoccupied energies with an important DOS reduction around $\varepsilon_F$ of a factor close to ten.

To investigate the electronic properties of the ground state and the mechanism beyond the gap opening we measured the temperature-dependent differential conductance $dI/dV$ as presented in Fig.~\ref{comp_dI_dV}b. Between 40~K and 30~K, the lowest temperature range where ARPES measurements were performed \cite{ncomms2617,Jaeger2018}, the LDOS at $\varepsilon_F$ is strongly reduced but still finite. Below $\sim30$~K, this depletion increases strongly in the energy range [-400;+500]~meV and within our experimental uncertainty, the tunneling conductance reaches zero near 25~K. The spectral features shift monotonically with temperature: occupied-state peaks move to higher binding energy while unoccupied features shift to larger energies. From our low-temperature measurements, we attribute earlier reports of a small gap at 4 K \cite{Modesti2007,Odobescu2017} to either surface contamination or insufficient structural cleanliness that we discuss in Appendix~\ref{APPENDIX_experiment}.

At 4~K the surface displays a fully developed large insulating gap, with a vanishing conductance in the  [-400;+500]~meV energy range. However, this suppression of spectral weight is overestimated by tip-induced band bending due to the poor surface conductivity leading to a temperature-dependent and energy-dependent peak shifts $\delta (E,T)$ of the measured STS spectra below 40~K. These shifts are smaller (larger) for negative (positive) energies and renormalize the measured band gap. Using the T-dependence presented in Fig.~\ref{comp_dI_dV}b and the identification of the main Si and Sn peaks presented in Fig.~\ref{comp_dI_dV}e, we could provide an experimental estimate of the insulating gap at 4~K : $\Delta=$440~meV $\pm$ 120~meV (see Appendix \ref{APPENDIX_comparison} and Ref.\cite{PhysRevB.73.161302} for details).

This value greatly exceeds previous experimental estimates reported in Ref.~\cite{Modesti2007,Odobescu2017,Ming2017} or theoretical prediction either using many body approaches based on the Hubbard model \cite{ncomms2617,Hansmann2013a} or DFT+U calculations  \cite{PhysRevLett.98.086401,Iannetti_2025} (such studies were conducted over a range of U parameters, U$=0$ to 4~eV). 
In addition, Fig.~\ref{comp_dI_dV}c,g shows that we can exclude spurious surface-related doping effects on the measured STS spectra by comparing the spatially-averaged LDOS with the local LDOS variations encountered at typical well-identified defects sites (representing $\sim$2.5\% of Sn sites), i.e. Si-substitution, Sn-vacancy and dopant-substitution. We observe small spectral shifts at Si or Sn peaks but no significant local modification of the chemical potential with the zero bias reference located close to the middle of the gap (taking into account the larger tip-induced band bending for $E>0$ than for $E<0$), which fully supports the $\sqrt{3}$-Sn phase being nominally undoped at 4~K.


Remarkably, despite being very large ($440$~meV/k$_B \simeq 5300$~K), the closing of the gap occurs at very low temperature (around $25$~K) with no evident structural modification of the surface from our STM analysis, while one would expect such a large energy gap to remain fully developed at 300~K and above. A plausible explanation of such apparent contradiction lays in the (arsenic) n-doped Si substrate used in the experiments. The As donor level is located very close in energy to the unoccupied Sn surface state (As binding energy is about 50~meV) and thermal excitations from the n-doped substrate to the Sn surface band could occur even at low temperature. Although further studies are needed to confirm this hypothesis, we attribute the low-temperature melting of the Mott insulating state to this self-doping mechanism.



\section{Theory: modelling the structural and electronic properties of the correlated 1/3 monolayer $\sqrt{3}\times\sqrt{3}$~R30$^{\circ}$-Sn/Si(111)}\label{sec_th}

\begin{figure}
\centering 
\includegraphics[width=\linewidth]{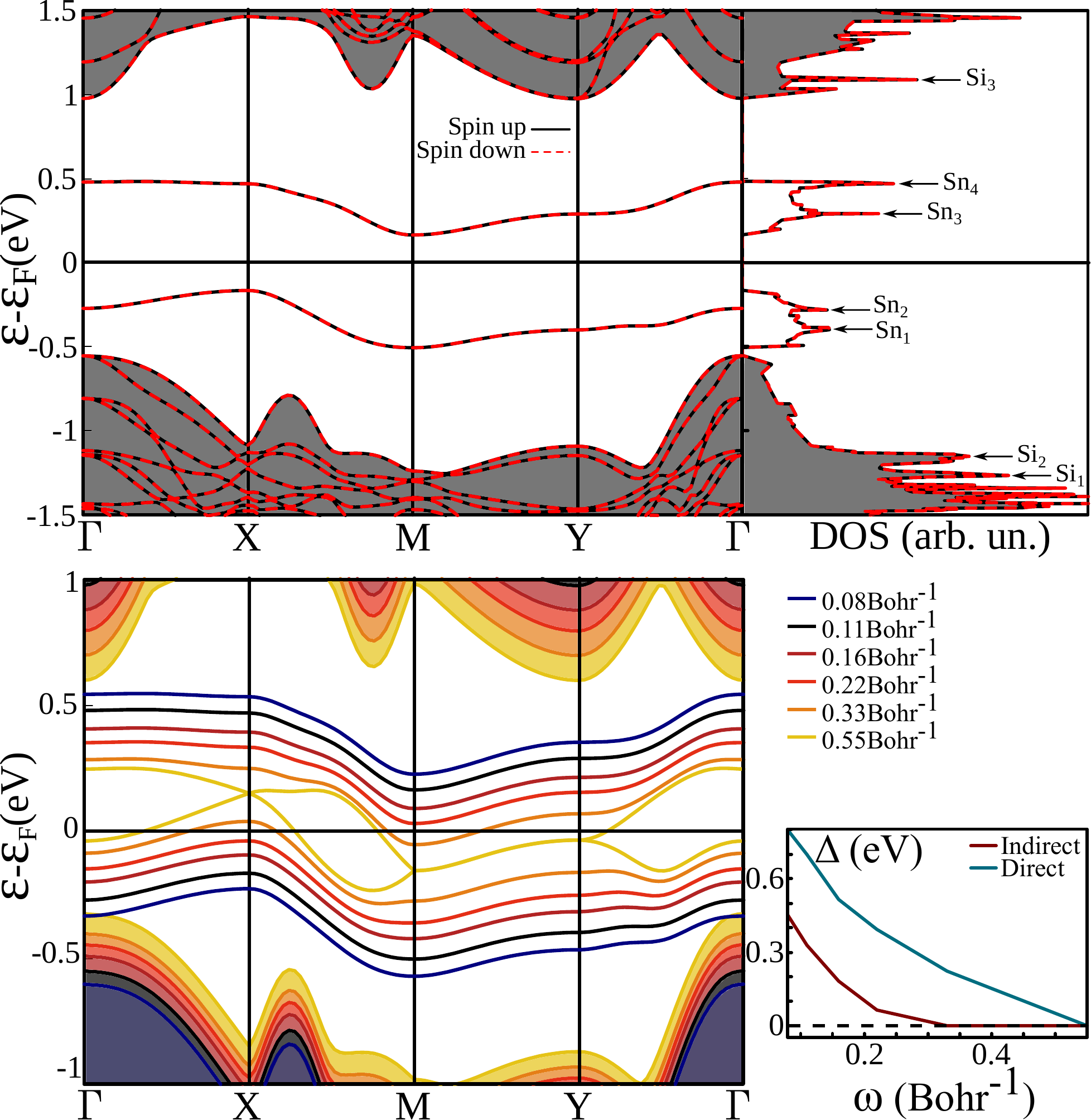}
\caption{Top panel: electronic band structures for the antiferromagnetic $2\sqrt{3}\times\sqrt{3}$~R30$^\circ$-Sn/Si(111) system along the $\Gamma-X-M-Y-\Gamma$ path of the Brillouin zone (see Fig.\ref{topos_structure}) along with the corresponding density of states. Marked peaks in the density of states correspond to the measured ones (see Fig.\ref{comp_dI_dV}). The projected Si bulk band structure is indicated as gray-shaded areas. Bottom left panel: electronic band structures, same as top panel, as a function of the HSE screening parameter $\omega$, from $0.08$ (blue lines) to $0.55$ Bohr$^{-1}$ (ocher lines) (see text). Bottom right panel: direct (at $M$ point) and indirect band gap as a function of the screening parameter. Note that $\omega=0.11$ Bohr$^{-1}$ is the standard HSE06 value, used in the upper panel.}\label{el}
\end{figure}

To clarify the origin of the large-gap insulating phase and to determine which are the relevant mechanisms responsible for stabilizing it, we studied the system from first principles.
We carried out density functional theory (DFT) calculations of the Sn/Si$(111)$ surface using the hybrid HSE06 functional \cite{doi:10.1063/1.1564060,doi:10.1063/1.2204597} (see Appendix \ref{APPENDIX_theory} for details).
We first computed the total energy for different ferromagnetic and antiferromagnetic configurations in the $2\sqrt{3}\times\sqrt{3}$~R30$^\circ$ Sn/Si(111) cell inferring that the collinear antiferromagnetic (AFM) ordering is the ground state with an energy gain of $\sim 0.07$~eV/Sn with respect to the non-magnetic state.
As shown in the top panel of Fig.\ref{el}, the collinear AFM ground state is insulating and present a band gap of $\Delta \simeq 0.35$~eV. 
Our theoretical gap is remarkably larger than that computed in the literature by means of DFT+U calculations \cite{Lee2014,Santoro,Li2013,Li2011,PhysRevLett.98.086401,PhysRevB.62.1556,Zahedifar_2019,Marchetti_2024,Craco_2024}, highlighting the importance of considering exact exchange in the Kohn-Sham Hamiltoninan, given the experimentally measured large gap.
The computed band gap ($\Delta \simeq$0.35~eV) is in good agreement with the STS measurements ($\Delta$=0.44$\pm 0.12$~eV), considering the tip induced band bending correction (see Appendix\ref{app_STS}).


Understanding the role of the non-local exchange energy contribution 
in the realization of such large gap 
is nontrivial and reflects the nature of the $\sqrt{3}$-Sn system and its related compounds. Indeed, the Sn-p$_z$ wavefunction significantly extends into the substrate, involving multiple Si atoms both laterally and in depth, as shown in Fig.\ref{topos_structure}f. 
To quantify the importance of non-local interaction effects, we conducted a theoretical experiment by calculating the band structure of the Sn/Si(111) AFM state as a function of the HSE screening parameter $\omega$, 
 which controls the real-space cutoff of the exact-exchange correction to the PBE functional \cite{doi:10.1063/1.1564060,doi:10.1063/1.2204597}, (Fig.\ref{el}, bottom panels). 
We evince that from large (local in space) to small (spatially extended) $\omega$ values, the Sn band  structure goes from a purely non-magnetic metallic state ($\omega=0.55$~Bohr$^{-1}$) to a large-gap antiferromagnetic system ($\omega=0.08$~Bohr$^{-1}$). 
The pronounced dependence of $\Delta$ on the screening parameter $\omega$ reveals 
the importance of the extended nature of the Sn p$_z$ orbitals, stretching into the Si substrate. This promotes the substrate itself as a key mediator of non-local interactions, by building the intersite correlations that help to stabilize the large-gap antiferromagnetic state. On the other hand, the DFT+U approach struggles to describe such a non-locality, giving a smaller gap value of about 0.2~eV as we show from a detailed comparison between HSE06 and DFT+U (see Fig.~\ref{fig_SM:DFT} and Appendix~\ref{APPENDIX_theory} for further discussions). These findings lead to a low-energy description of $\sqrt{3}$-Sn (e.g., in a frame of an extended one-band Hubbard model) that contains non-local interaction terms in the Hamiltonian.


\section{Quasiparticle interferences and signatures of the antiferromagnetic order}\label{mag_part}

\begin{figure*}
\centering 
\includegraphics[width=\linewidth]{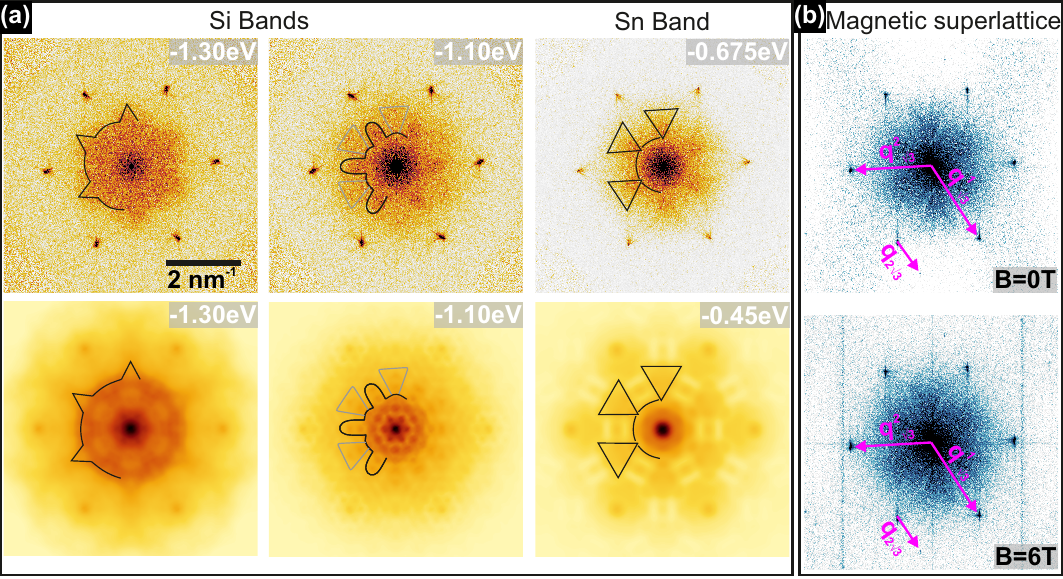}
\caption{(color online) Revealing signatures of the Sn, Si bandstructures and magnetic supercell. a) Energy dependent quasiparticle interferences (QPI) maps (black and white). The top row represents experimental QPI acquired under $B=0$ magnetic field while the bottom raw gives the theoretical calculated ones. Each top panel represents the numerical Fourier transform of a $dI/dV(E=eV)$ map acquired at the indicated energy at 4~K in the same $50\times50$~nm$^2$ area of a single $\sqrt{3}\times\sqrt{3}$~R30$^{\circ}$-Sn/Si(111) domain. The bottom row reports symmetrized joint density of states calculations obtained using the \textit{ab-initio} electronic band structure shown in Fig.\ref{el}, top panel. The colored drawings are guide for the eyes. b) Experimental QPI maps acquired at $B=0$ and $B=6$~T perpendicular magnetic field. In addition to the usual Bragg peaks indicated by $q^1_{\sqrt{3}}$ and $q^2_{\sqrt{3}}$, sharp superlattice spots of weaker intensities are seen, indexed by the vector $q_{2\sqrt{3}}$, corresponding to the colinear antiferromagnetic ordering of the spin $1/2$ lattice.}\label{QPI_mag_order}
\end{figure*}

An elegant and powerful way to infer information about order parameter symmetry, topology, spin and band-structure properties is to use quasiparticle interferences (QPI) STS measurements. QPI are a phase-coherent manifestation of the scattering of elementary electron- or hole-excitations by defects \cite{SimonFTSTS,Chen2017,Lin2020}. QPI measurements are obtained by performing the numerical Fourier transform of a $dI/dV(E=eV,x,y)$ map measured for an energy $E=eV$ associated to the applied bias voltage $V$ on a real space ($x,y$) grid. In the closely related $\sqrt{3}\times\sqrt{3}$~R30$^\circ$-Pb/Si(111) system we showed that this technique can be successfully used to map out the main elastic scattering vectors of the Fermi surface and hence reveal the underlying Rashba spin-orbit coupling and shape/size of the Fermi surface of this correlated 2D metal presenting a combined CDW/charge order \cite{Tresca2018}. Nevertheless, as discussed in recent works \cite{Tresca2023,PhysRevB.110.085109}, such measurements must be performed with particular care when the LDOS varies strongly within the unit cell, as often happens in correlated or charge-ordered systems 
(see Appendix \ref{APPENDIX_experiment} for more details).

In Fig.~\ref{QPI_mag_order}a, the top panel present the experimental QPI maps acquired at 4~K in a single structural $\sqrt{3}$-Sn domain (see also Appendix \ref{APPENDIX_experiment}), associated to the $dIdV$ maps presented in Fig.~\ref{comp_dI_dV}. The bottom panels present the calculated QPI corresponding to the symmetrized joint density of states, obtained using the \textit{ab initio} electronic band structure shown in the top panel of Fig.\ref{el} (see Appendix \ref{APPENDIX_theory}). The energy correspondence between experimental and theoretical QPI maps was determined from the comparison between the experimental and theoretical spectral features present in the LDOS. Four main peaks labelled Sn$_1$, Sn$_2$, Sn$_3$, Sn$_4$ can be identified in the Sn p$_z$ band and 3 peaks labelled Si$_1$, Si$_2$, Si$_3$ in the Si states over the [-1.8;+1.8]~eV probed energy range (see Fig.\ref{comp_dI_dV}g and top panel of Fig.\ref{el}). Following Ref.~\cite{PhysRevB.73.161302}, our analysis shows that there is a systematic difference between measured and calculated peak energies that can be quantitatively inferred and that strongly depends on the bias voltage polarity as already anticipated.

The detailed comparison between the measured and calculated QPI maps in Fig.\ref{QPI_mag_order} validates our combined experimental and theoretical investigation of the ground-state electronic structure. The Bragg peaks associated to the structural $\sqrt{3}$-Sn unit cell are visible as sharp spots for all experimental maps (labeled by ${\bf q^{1,2}_{\sqrt{3}}}$ in Fig.\ref{QPI_mag_order}b). Deep in the occupied states, the Si band is signaled by a star-shaped feature oriented along the $K$ points, associated to the broad peak Si$_1$ located both experimentally and theoretically around -1.3~eV. The star-shaped calculated QPI is clearly visible with a comparable $k$-space extension, showing good agreement (see Fig.\ref{QPI_mag_order}a). Moving to the Si$_2$ peak at $\sim -1.1$~eV, the star-shaped features around the $K$ points evolve to stripe-like signals directed between the $\Gamma$ and $M$ points, both in the measured and calculated spectra. QPI maps for the Sn band is presented for $-0.675$~eV. This occupied state energy is associated to the local LDOS maxima Sn$_1$ (experimentally near $-0.65$~eV, theoretically near $-0.45$~eV). The patterns are complex and several features can be recognized at similar locations between the measurement and the calculation, looking overall as flower-shape patterns drawn as triangles pointing toward the internal brighter corona.
As we mentioned above, at 4~K the tip-induced band bending affects much more unoccupied states than occupied states \cite{PhysRevB.73.161302}. For this reason we mostly concentrated on $E<0$ for the QPI analysis.

A closer inspection of the QPI maps reveals additional sharp and weaker Bragg-like spots (indicated by ${\bf q_{2\sqrt{3}}}$ in Fig.~\ref{QPI_mag_order}b). These satellite features do not disperse in energy and are visible in various QPI maps for Sn and Si band at $E>0$ and $E<0$ energies (see the movie indicated in the appendix~\ref{APPENDIX_experiment}). They also remain under a 6~T perpendicular magnetic field. The experimentally extracted values are ${\bf q_{2\sqrt{3}}}=(0.48 \pm{0.2}) {\bf q^{1}_{\sqrt{3}}}$ at 0~T and ${\bf q_{2\sqrt{3}}}=(0.49 \pm{0.2}) {\bf q^{1}_{\sqrt{3}}}$ at 6~T. They are consistent with a $2\sqrt{3}\times\sqrt{3}$~R30$^\circ$ real-space supercell associated with a single magnetic domain, as predicted for the collinear antiferromagnetic ground state. We therefore interpret our combined theoretical and experimental analysis as the first direct experimental proof of the existence of a row-wise $2\sqrt{3}\times\sqrt{3}$~R30$^\circ$ antiferromagnetic ordering of the spin-$1/2$ triangular lattice in undoped $\alpha$-Sn/Si(111).

\section{Conclusion}

This study provides a thorough elucidation of the long-debated nature of the $\sqrt{3}$-Sn ground state at low temperature, which for decades has been regarded as one of the most promising realizations of the two-dimensional Mott-Hubbard model on a triangular lattice. By combining low-temperature spectroscopy and quasiparticle interference mapping with first-principles calculations, we establish that below 30 K the $\sqrt{3}$-Sn surface undergoes a pronounced transition into a fully developed insulating phase characterized by the opening of a large energy gap of approximately 440 meV at 4 K, enlarged by a tip-induced band bending mechanism. This value, being five to ten times larger than previously proposed by both experiments and theory, redefines the electronic landscape of this system, rationalized by its theoretical description.

In particular, the unusually large gap and its sensitivity to the presence of the substrate highlight the limitations of models relying solely on local Coulomb repulsion. Our results indicate that non-local contributions to correlations, mediated by the silicon substrate, play a decisive role in stabilizing the large-gap insulating phase. This insight has important implications for theoretical modeling: any realistic low-energy description of $\sqrt{3}$-Sn (and the related compounds) must incorporate extended Coulomb terms and intersite exchange processes with the substrate. 

The coexistence of such a large gap with a relatively low metallization temperature of about 25~K appears at first glance inconsistent with conventional expectations for a large-gap Mott insulating phase. However, this apparent discrepancy can be reconciled by considering a thermally driven self-doping mechanism. In this scenario, donor impurity states associated with arsenic dopants are thermally activated and located energetically above or close to the Sn p$_z$ upper Hubbard band \cite{Mizokawa_PRB_selfoding,Wang_ArXiv_selfdoping,Rosciszewski_PRB_selfdoping,Craco_PRB_selfdoping,Kanno_PRB_selfdoping}, thereby dynamically altering the carrier concentration as the temperature rises. 
This self-doping picture not only offers a natural explanation for the reduced transition temperature and metalized STS spectra reported above 77~K but also underscores the subtle interplay between substrate doping and electronic correlations in shaping the ground state. However, despite being plausible, these conjectures merit future investigations that could infer the underlying metallization mechanism, possibly controlling the transition temperature.

Beyond the insulating behavior, our quasiparticle interference measurements reveal distinct fingerprints of a collinear antiferromagnetic order, emerging in single structural domains and giving rise to signatures of a non-dispersive signal with a 2$\sqrt{3}\times\sqrt{3}$~R30$^\circ$ periodicity, unchanged under a 6~T perpendicular magnetic field. The corresponding real-space reconstruction, consistent with a 2$\sqrt{3}\times\sqrt{3}$~R30$^\circ$ magnetic supercell, finds strong theoretical backing from DFT calculations. The good agreement between the experimental and theoretical QPIs 
further supports a non-local interaction picture as the key ingredient in understanding correlation effects in $\sqrt{3}$-Sn.


\section{Acknowledgements}
We acknowledge CINECA (ISCRA initiative) for computing resources. This work was supported by French state funds managed by the ANR within the Investissements d'Avenir programme under reference  ANR-11-IDEX-0004-02, and more specifically within the framework of the Cluster of Excellence MATISSE led by Sorbonne Universit\'es and ANR contract RODESIS ANR-16-CE30-0011-01. M. I., C. T. and G. P. acknowledge the Laboratori Nazionali del Gran Sasso for computational resources.
C. T. acknowledges financial support from the National Recovery and Resilience Plan (NRRP), Mission 4, Component 2, Investment 1.1, funded by the European Union – NextGenerationEU, for the projects “DAREDEVIL” (CUP D53D23002960001) and “SHEEP” (CUP B53D23028580001). This work has also been funded by the European Union - NextGenerationEU, Mission 4, Component 2, under the Italian Ministry of University and Research (MUR) National Innovation Ecosystem grant ECS00000041 - VITALITY - CUP B43C22000470005, C. T. acknowledges the Università degli Studi di Perugia for the support with this project.

\appendix
\section{Experimental details}
\label{APPENDIX_experiment}

\subsection{Sample growth}

\begin{figure*}
    \centering
    \includegraphics[width=\linewidth]{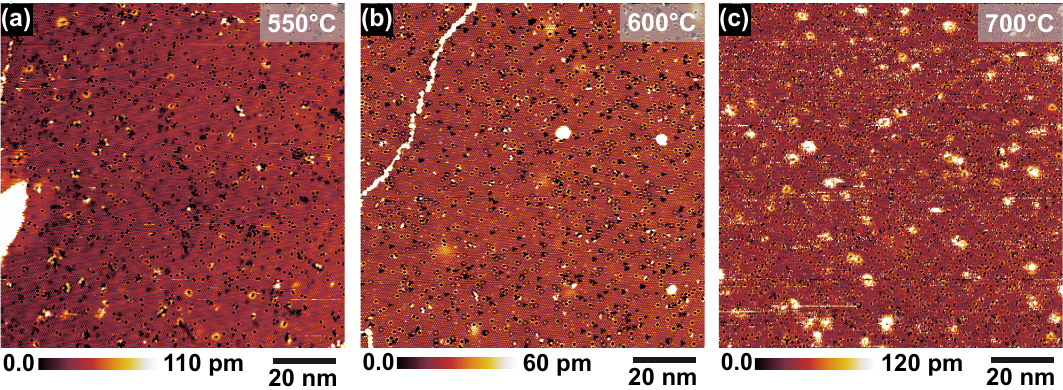}
    \caption{\textbf{Optimizing the temperature growth: defects density}. STM topographies of three different samples showing atomic resolution of a $100\times100$~nm$^2$ area of $\sqrt{3}\times\sqrt{3}~R30^{\circ}$-Sn/Si(111) measured at $T=4.2$~K with $V_{bias}=-2.0~$V and  $I=50~$pA for 0.33~monolayer of Sn deposited on a n-doped $7\times7$-Si(111) substrate held at: a) 550$^{\circ}$C, b) 600$^{\circ}$C, c) 700$^{\circ}$C. The lowest point defects density is found at 600$^{\circ}$C.}
    \label{fig:SM_optim_temp_defects}
\end{figure*}

\begin{figure*}
    \centering
    \includegraphics[width=\linewidth]{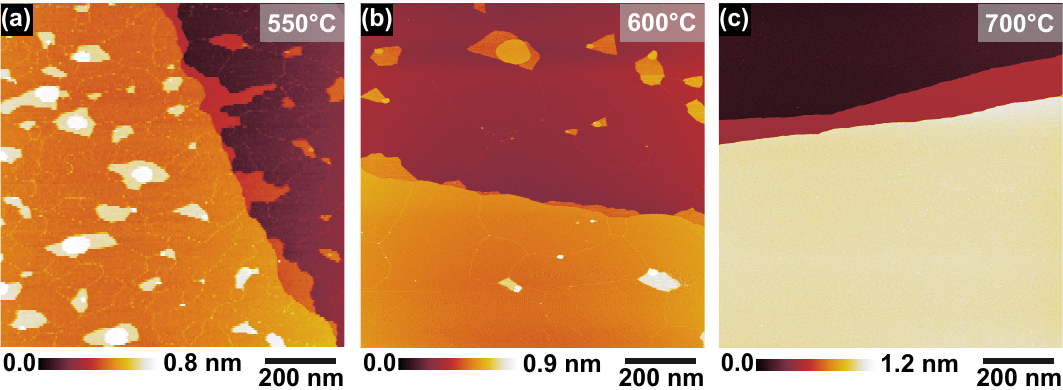}
    \caption{\textbf{Optimizing the temperature growth: domain size}. Large scale STM topographies of three different samples showing a $1\times1~\mu$m$^2$ area of $\sqrt{3}\times\sqrt{3}~R30^{\circ}$-Sn/Si(111), measured at $T=4.2$~K with $V_{bias}=-2.0~$V and  $I=50~$pA, for 0.33~monolayer of Sn deposited on a n-doped $7\times7$-Si(111) substrate held at: a) 550$^{\circ}$C, b) 600$^{\circ}$C, c) 700$^{\circ}$C. Large single domain size above 100~nm is obtained above 600$^{\circ}$C.}
    \label{fig:SM_optim_temp_size}
\end{figure*}

\begin{figure}
    \centering
    \includegraphics[width=\columnwidth]{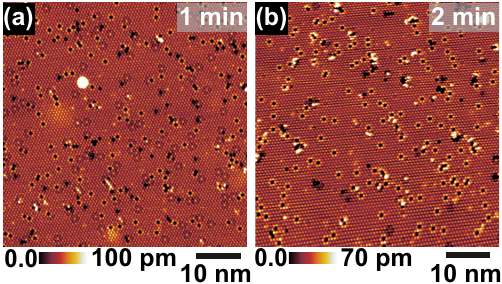}
    \caption{\textbf{Optimizing the temperature growth: annealing time}. STM topographies of two  different samples showing atomic resolution of a $50\times50$~nm$^2$ area of $\sqrt{3}\times\sqrt{3}~R30^{\circ}$-Sn/Si(111) measured at $T=4.2$~K with $V_{bias}=-2.0~$V and  $I=50~$pA for 0.33~monolayer of Sn deposited on a n-doped $7\times7$-Si(111) substrate held at 600$^{\circ}$C and kept at 600$^{\circ}$C for an annealing time of: a) 1~min, b) 2~min. More point defects occur above 2 minutes than for 1 minute annealing time.}
    \label{fig:SM_optim_anneal_time}
\end{figure}

 The $\sqrt{3}$-Sn single-layer phase was prepared using a well-known procedure \cite{ncomms2617,Ming2017,Odobescu2017}. We used heavily arsenic $n$-doped Si(111) substrates ($N_{As}\simeq 1.0$  $10^{19}\cdot$cm$^{-3}$) having temperature resistivity in the m$\Omega\cdot$cm range. The Si(111)-$7\times7$ reconstruction was prepared using repeated flashes around 1100$^{\circ}$C followed by a slow annealing down to about 600$^{\circ}$C. The substrate temperature was then stabilized at 600$^{\circ}$C and 0.33 monolayer of Sn was evaporated in about two minutes from a triple e-beam evaporator operated with a high-voltage ions suppressor, followed by an annealing at 600$^{\circ}$C for about one minute. We determined this optimum temperature growth of 600$^{\circ}$C (see Fig.\ref{fig:SM_optim_temp_defects} and Fig.\ref{fig:SM_optim_temp_size}) and annealing time (see Fig.\ref{fig:SM_optim_anneal_time}) taking into account the following simultaneous constraints: large $\sqrt{3}$-Sn single domains (larger than 50$\times$50~nm$^2$) in order to perform STS spectroscopy maps for QPI experiments (see Fig.\ref{fig:SM_optim_temp_size}) and small density of point defects (see Fig.\ref{fig:SM_optim_temp_defects}). These 3 figures illustrate our conclusions and choices made in order to optimize the $\sqrt{3}$-Sn sample surface quality.
 
 As described above, we took a lot of care in order to produce the cleanest $\sqrt{3}$-Sn main phase as possible. In addition we also took care of improving the direct electrical contact to the sample surface in order to facilitate electronic charge injection and reduce possible band bending effects. The bias voltage electrical contacts are provided by thin molybdenum plates pressed on top of the sample surface. Various attempts to improve the surface-top Mo electrode electrical contact using a pre-deposited Tantalum layer or a Pt wire resulted in a degraded surface quality characterized by an increased defects density, a reduced energy gap as shown below in the next section, or even sometimes rendering the sample metallic at 4~K. Thus, our detailed sample preparation study suggests that some discrepancies existing between various published spectroscopic results may be associated to the different cleanliness procedures used in the surface preparation and to the practical way electrical contacts to the sample surface are effectively realized. In our case, the cleanest samples as characterized from STM/STS experiments were always associated to large-gap insulating samples at $T=4$~K, as exemplified below in the next section. We note that in other families of well studied Mott insulators like iridates for example, smaller energy gaps associated to less clean samples have been also reported while the well-prepared ones show larger gaps \cite{srep03073,PhysRevB.89.085125,PhysRevX.5.041018,nphys3894}.

\subsection{STS measurements}\label{app_STS}
The STM/STS experiments were carried out using a homemade 300~mK apparatus based on a $^3$He single-shot cryostat \cite{Serrier2013,Brun2014,Brun2017}. For the need of the present work, our STM was operated at 4~K using He exchange gas. We used mechanically cut metallic PtIr tips. The $dI/dV(V)$ spectra were obtained by numerical derivation of the raw single $I(V)$ curves. The individual $I(V)$ spectra were convolved with a Gaussian filter compatible with the thermal broadening at 4~K. The negative (positive) bias voltage corresponds to occupied (empty) sample states.

We have carried out STS spectroscopy measurements on $\sqrt{3}$-Sn single domains having a lateral size larger than 50-100~nm. For such domain sizes the spectral characteristics presented hereafter are well established and reproducible. The $I(V)$ spectra were measured far enough from boundaries with neighboring 2$\sqrt{3}\times 2\sqrt{3}$-Sn domains in order to probe intrinsic electronic properties.

We defined and extracted the energy gap $\Delta$ from the energy range around the Fermi energy where the measured $dI/dV$ signal reaches zero, i.e. the noise level of the preamplifier. However below 40~K, we noticed that a strong tip induced band bending (TIBB) develops shifting in energy the observed spectral features associated to the Sn and Si bands, an effect that is much less pronounced for occupied states ($E<0$) than unoccupied ones (see Fig.~\ref{comp_dI_dV}). Such an effect was already reported for the bare Si(111) surface (see \cite{PhysRevB.73.161302} and references therein). In order to provide a reliable estimate of $\Delta$ we first estimated the TIBB for $E<0$ by subtracting the Si$_1$ and Si$_2$ peak energies at 40~K and at 4~K, taking into account the spatial variations measured at 4~K from Fig.~\ref{comp_dI_dV}e. This led to TIBB $\simeq -0.2 \pm 0.04$~eV. Half the gap value $\Delta /2$ for $E<0$ was then extracted from the Sn band onset, the latter being determined from detailed QPI analysis (average energy between the appearance of diffuse scattering and Bragg peaks in QPI map): $E_{Sn-onset} = -0.42 \pm 0.02$~eV, as $\Delta /2 = E_{Sn-onset} -$TIBB. This leads to $\Delta = 0.44 \pm 0.12$~eV, corresponding to the value given in the main text. This gap value, as discussed in the main text, does not depend on the samples as soon as their quality and electrodes electrical contacts were clean and did not bring any extra surface contamination. Please note that the value given here for the occupied states Sn band onset is in very good agreement with an independent study conducted recently analyzing the statistical LDOS distribution of the extended Sn/Si(111) states \cite{Lizee2025}.

\begin{figure*}
    \centering
    \includegraphics[width=\linewidth]{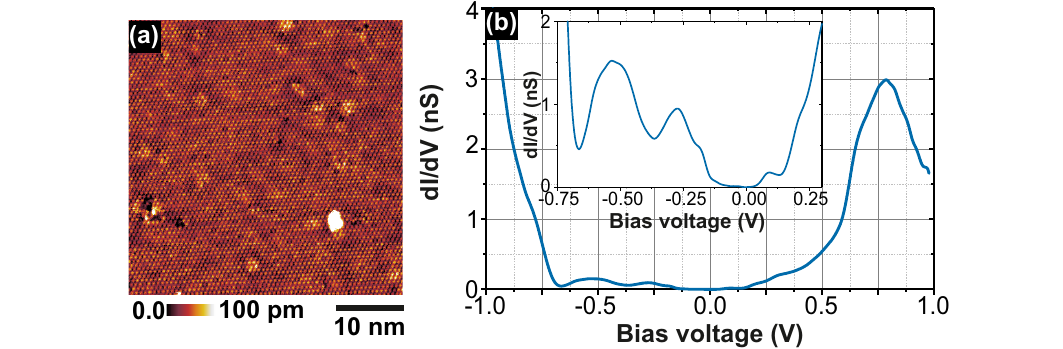}
    \caption{a) STM topography with atomic resolution measured in a $100\times100$~nm$^2$ area at $T=4.2$~K in $\sqrt{3}\times\sqrt{3}~R30^{\circ}$-Sn/Si(111) with $V_{bias}=-1.0~$V and $I=50~$pA. A Ta contact layer was predeposited onto the oxidized Si(111) surface prior to Si(111) preparation in order to improve the surface electrical contact with top Mo electrodes. This results in a surface contamination where the observed defects are different than the usual ones presented in Fig.\ref{topos_structure} or \ref{comp_dI_dV}. b) Average $dI/dV$ spectroscopy measured over the whole area. A much smaller energy gap is measured on this contaminated sample than the one presented in clean samples presented in Fig.\ref{comp_dI_dV}.}
    \label{fig:SM_contam}
\end{figure*}

An example of contaminated sample surface prepared using a predeposited Ta contact layer close to one of the Mo surface electrode is shown in Fig.\ref{fig:SM_contam}a. In this case the Ta layer was deposited onto oxidized Si surface in cleanroom facility prior to the first Si(111) surface preparation in UHV. Fig.\ref{fig:SM_contam}a shows that the defects observed by STM are different than the usual ones characteristic of a clean surface presented in Fig.~1 of the main text or in Fig.\ref{comp_dI_dV}c. In such surface contaminated cases, as also using Pt wires to improve the surface electrical contacts, the energy gap measured by STS at 4~K presented in ~\ref{fig:SM_contam}b was found to be much smaller than found in clean samples presented in Fig.\ref{comp_dI_dV}. The much smaller gap value found in contaminated surfaces is actually comparable to the one reported by Modesti \textit{et al.} and Odobescu \textit{et al.} Ref.~\cite{Modesti2007,Odobescu2017}. This situation reminds us the case of other Mott physics materials like iridates where smaller gap values were also reported depending on the surface preparation procedure while larger gaps are associated to cleaner samples \cite{srep03073,PhysRevB.89.085125,PhysRevX.5.041018,nphys3894}.

The QPI experiments presented in Fig.~\ref{QPI_mag_order} were obtained from the numerical Fourier transforms of $dI/dV(E,x,y)$ maps measured from a dense $256\times256$ grid acquired over a $50\times50$~nm$^2$ area presented in Fig.~\ref{comp_dI_dV}. For the panel (a) the spectroscopy set-point was $V_{bias}=-2$~V and $I=200$~pA with 2200 points acquired in [-2;+2]~V range. The QPIs in panel (b) were acquired from two additional separate measurements with a similar $256\times256$ grid but using a smaller [-1;0]~V range in order to increase the energy resolution in the $E<0$ Sn band, at 0~T (top panel) and 6~T (bottom panel). In panel (a) the experimental QPI were symmetrized assuming two mirror symmetries in order to compare with the symmetrized theoretical ones, which enhances the signal-to-noise ratio. In panel (b) the QPI maps were not symmetrized in order to provide the most precise analysis of the position of the additional satellite peaks attributed to the magnetic 2$\sqrt{3}\times\sqrt{3}$~R30$^\circ$ period.

\section{Theoretical details}
\label{APPENDIX_theory}
\subsection{\textit{Ab initio} calculations}
\label{app_theory_abinitio}

We model the Sn/Si(111) surfaces by considering a layer of Sn atoms on top of 5 Si bilayers orthogonal to the (111) Si-bulk direction. The bottom dangling bonds are capped with hydrogen atoms fixed to the relaxed positions obtained by capping one side of the pristine Si(111) surface.
The atomic positions of the first four Si-substrate bilayers below the Sn single layer are optimized whereas the remaining fifth bilayer is fixed to the Si bulk positions. The periodic image of the system was placed at $\sim$20~\AA\ in the out-of-plane direction.

In order to study the collinear antiferromagnetic phase for Sn/Si(111), we considered a surface supercell containing 2 Sn atoms: the 2$\sqrt{3}\times\sqrt{3}$~R30$^\circ$ cell with respect to the Si-(1$\times$1) surface reconstruction.

Density functional theory (DFT) calculations using the hybrid-functional HSE06\cite{doi:10.1063/1.1564060,doi:10.1063/1.2204597} are performed with \textsc{FHI-AIMS}\cite{FHI-AIMS_2009,FHI-AIMS_2012} package and with localized basis sets.
Integration over BZ was performed on a 6$\times$12$\times$1 k-point grid and a Gaussian smearing of 0.01~eV was adopted.
In this framework, we optimized the internal coordinates.
DFT+U \cite{PBE_1996,PBE_2008} calculations are instead performed as implemented in \textsc{Quantum-Espresso} \cite{QE_2009,QE_2017,QE_2020}. The Hubbard correction to the density functional is computed using the linear response theory described by Cococcioni \textit{et al.} \cite{Cococcioni_2005} (see Appendix \ref{app_theory_nonlocal}). DFT (PBE) calculations with \textsc{Quantum-Espresso} were carried out modeling the Si(111) substrate similarly to the HSE06 calculation but considering only 3 Si bilayer, to fairly reproduce the silicon bulk gap\cite{Tresca2021}. An energy cut-off of 45~Ry was adopted for the wavefunction expansion while the integration grid over the BZ is unchanged with respect to the HSE06 calculations.

\subsection{Theoretical QPI maps}
\label{app_theory_qpi}

The theoretical QPI maps were computed within the Joint Density of States
(JDOS) approximation as described in Ref. \cite{Tresca2023} where the JDOS is defined as:
\[
JDOS\left(\boldsymbol{q},\omega\right)=\sum_{\boldsymbol{k}}\sum_{n,m}\vert M_{n\boldsymbol{k},m\boldsymbol{k}+\boldsymbol{q}}\vert^{2}A_{n\boldsymbol{k}}\left(\omega\right)A_{m\boldsymbol{k}+\boldsymbol{q}}\left(\omega\right),
\]
where $M_{n\boldsymbol{k},m\boldsymbol{k'}}=\left\langle \boldsymbol{k}n\left|Ve^{i\left(\boldsymbol{k}-\boldsymbol{k'}\right)\cdot\boldsymbol{r}}\right|\boldsymbol{k'}m\right\rangle $ is the scattering matrix element of a (non-magnetic) impurity with potential $V=V_{0}\sigma_{0}$ with $\sigma_{0}$ being the identity matrix in the spin space. $A_{n\boldsymbol{k}}\left(\omega\right)=\frac{W_{n\boldsymbol{k}}}{\omega-\epsilon_{nk}+i0^{+}}$ is the spectral function of the $n$-th band with momentum $\boldsymbol{k}$ at energy $\omega$ while $W_{n\boldsymbol{k}}$ is the unfolding spectral weight introduced because of the magnetic supercell involved and computed as implemented in the FHI-AIMS package. 
In order to take into account the $\boldsymbol{k}_{\parallel}=\boldsymbol{q}$ attenuation of STM experiments we multiply the $JDOS\left(\boldsymbol{q},\omega\right)$ by the suppressing factor \cite{PhysRevB.31.805}:
\[
JDOS\left(\boldsymbol{q},\omega\right)\rightarrow JDOS\left(\boldsymbol{q},\omega\right)e^{-K\left(\boldsymbol{q}\right)z}
\]
where $z\sim 5\text{Å}$ is a parameter representing the typical experimental distance between
the STM tip and the sample surface, and $K(\boldsymbol{q})=\sqrt{\frac{2m}{\hbar^{2}}\left(W-\epsilon_{\boldsymbol{q}}\right)+\boldsymbol{q}^{2}}$
with $W\sim4eV$ being the assumed work function, $\hbar$ the reduced Plank constant and $m$ the electron mass.
All computed QPI maps are symmetrized by 120$^\circ$ rotations due to the C$_3$ degeneracy of the antiferromagnetic stripes on the triangular lattice.
The reason behind this symmetrization lies in the peculiarity of the system under investigation and in the fragility of the long-range magnetic order due to intrinsic frustrations present in the crystal.
In fact, different phenomena could coexist, linked firstly to lattice frustration and secondly to the quantum nature of the system.
The first one is due to the small nearest-neighbor exchange constant (J$\simeq$ 10~meV) between different Sn atoms \cite{Iannetti2026} which should lead to a rather short-range magnetic order
This (eventually) causes the system to behave in a disordered way \cite{Goremychkin2008} or with very small antiferromagnetic domains.
Hence, the QPI map should be traced over all three orientations of the magnetic stripes with different relative weights, thus, giving an overall symmetrization of the signal.
The second phenomenon is related to the quantum nature of the system, which could lead to the coexistence of different magnetic orders (similarly to what occurs in the charge channel, as explained by Hansmann \textit{et al.}~\cite{Hansmann2016}).
A detailed analysis of these assumptions, however, is beyond the scope of this work and left for future investigations.


\subsection{Non-local nature of the insulating ground-state}
\label{app_theory_nonlocal}

\begin{figure}
    \centering
    \includegraphics[width=\columnwidth]{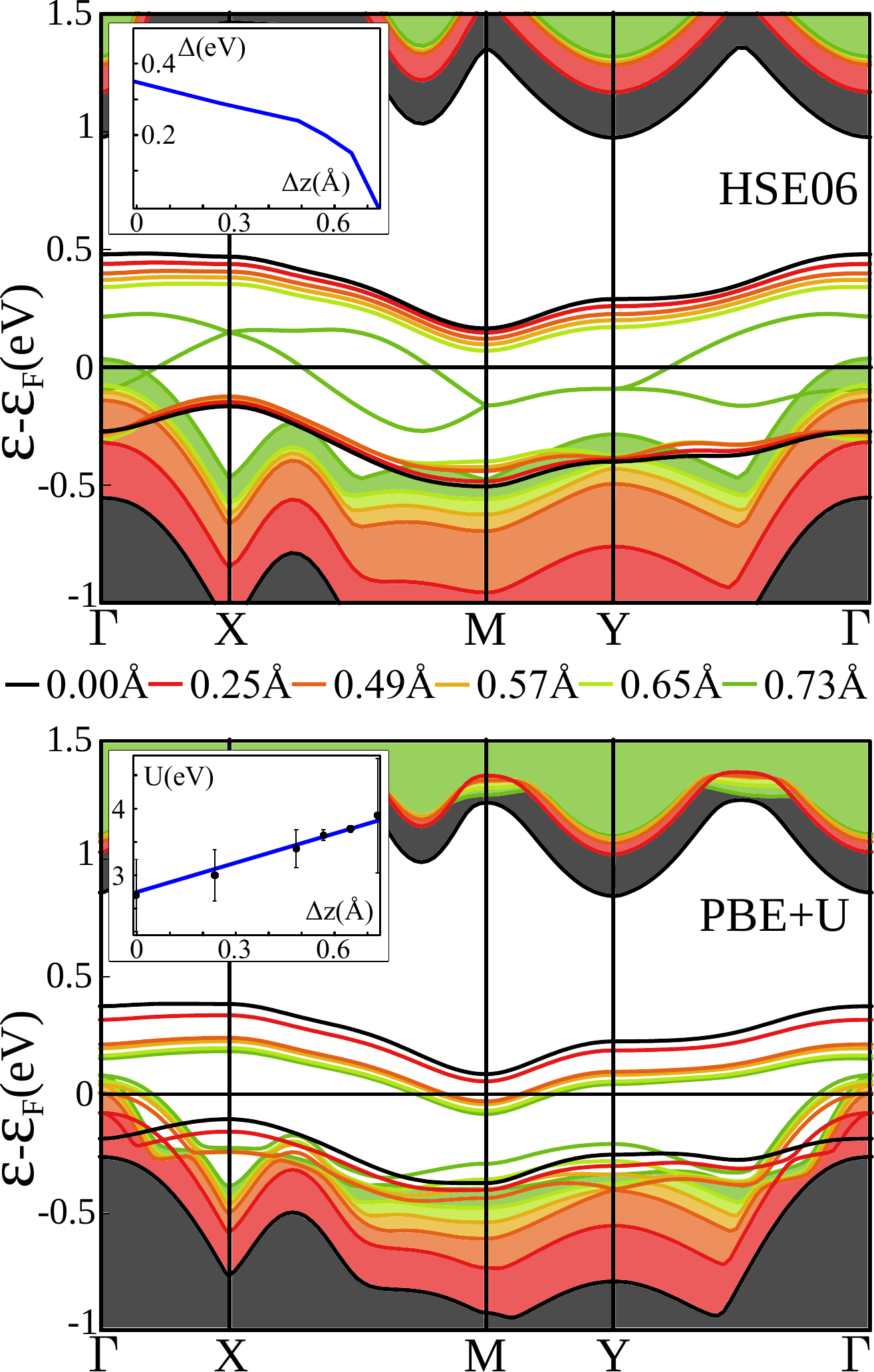}
    \caption{Electronic band structures for the antiferromagnetic Sn/Si(111)-$2\sqrt{3}\times\sqrt{3}$ system along the $\Gamma-X-M-Y-\Gamma$ path of the Brillouin zone as a function of the height of Sn adatoms in HSE06 (top panel) and PBE+U (bottom panel). The inset in the top (bottom) panel shows the value of the energy gap (the computed Hubbard U correction) as a function of $\Delta z$, i.e. the Sn height variation with respect to the equilibrium value. The $\Delta z$ values taken into account are reported in the legend.}
    \label{fig_SM:DFT}
\end{figure}

In Sec.~\ref{sec_th} we claim that the insulating magnetic ground state has a non-local nature, mediated by the substrate, as demonstrated by reducing the screening parameter $\omega$ in HSE calculations (see Fig.~\ref{el}). In order to strengthen our findings, we perform an additional computational experiment comparing HSE06 and PBE+U functionals. We raise the height of the Sn adatom $z$ from its equilibrium position $z_0$, hence reducing the overlap between the Sn wavefunctions and the substrate. The corresponding band structure is reported in Fig.~\ref{fig_SM:DFT}. In HSE06 calculation (top panel), the system abruptly loses its magnetic moment at $\Delta z = z - z_0 \simeq 0.69$~\AA, showing a non magnetic metallic band at the Fermi energy. On the contrary, in PBE+U calculations (bottom panel), the magnetic moment reduces but remains finite, even when the system becomes metallic, since the lower and upper Hubbard bands remain splitted. Moreover, the calculated Hubbard U correction increases as a function of $\Delta z$, revealing that the electronic screening due to the substrate reduces the Hubbard U correction to the PBE functional. This suggests that non-local correlations are crucial for realizing the insulating ground state. Therefore, care should be taken in describing this system using the \textit{ab initio} PBE+U functional or Hubbard-like Hamiltonians.\\
This clearly shows the key role played by the substrate as a mediator of the electronic correlations, highlighting the non-local nature of the electron-electron interaction of the Sn p$_z$ dangling orbital. 

\section{Comparison between experimental and theoretical spectral features: setting the proper correspondence}
\label{APPENDIX_comparison}
As discussed in the main text, the energy correspondence between experimental and theoretical QPI maps was carefully determined from the comparison between the experimental and theoretical spectral features present in the measured and calculated LDOS. 4 main peaks occur in the Sn bandstructure and 3 peaks in the Si bands over the [-1.8;+1.8]~eV experimental energy range. All the considered peaks for this correspondence are labelled in Figs.~\ref{comp_dI_dV}g and \ref{el} for the experimental and theoretical analysis, respectively. This analysis is reflected in the comparison of the QPI energies and shows that there is a systematic energy difference between the measured and theoretical bias. This phenomenon is well-known for gapped semiconductors studied by STS at low temperature and was thoroughly studied for the heavily n-doped Si(111) substrate used here~ (see \cite{PhysRevB.73.161302} and references therein). For Sn/Si(111) we expect this effect to be less strong than for pure heavily n-doped Si(111), since the Sn/Si(111) is a 2D metal at room temperature with a better surface conductivity than the one measured in the n-doped Si(111) substrate.

The energy difference existing between the true energy levels and the ones measured by STS at 4~K is induced by an additional surface potential drop related to band bending effect. The Si dopants (As atoms) are frozen out at 4~K and different regimes are encountered at the surface. There is an accumulation of charges for $E<0$, while a depletion is expected for $E>0$. In our case this energy difference is the smallest for the occupied states, where this effect is expected to be the smallest~\cite{PhysRevB.73.161302}. In contrast, for the unoccupied states, this energy difference is the largest~\cite{PhysRevB.73.161302}, and increases with energy. In this case this difference increases from about +0.5~eV at the Sn band onset to about +0.7~eV at the Sn band end.

Since band-bending effects are less important for occupied states \cite{PhysRevB.73.161302}, a proper energy normalization for all spectra can be found for $V_{bias}=2$~V and $I=200$~pA, which is deep into the Si bandstructure (about 1.1~eV). 
Then, we can exclude spurious surface-related doping effects on the measured STS spectra by comparing spatially-averaged LDOS variations around different typical well-identified defects sites (present in $\sim$2.5\% of Sn sites), i.e. Si-substitution, Sn-vacancy and dopant-substitution (see Fig.\ref{comp_dI_dV}c-e), observing no significant local modification of the chemical potential with the 0 bias reference located close to the middle of the gap.



\clearpage
\bibliography{bibliography}{}

\end{document}